\documentclass[useAMS,usenatbib]{mn2e}

\usepackage{graphicx}
\usepackage{amsmath}
\usepackage{amssymb}
\usepackage{color}
\usepackage{subfig}
\usepackage[breaklinks,colorlinks,citecolor=blue,linkcolor=red]{hyperref} 
\usepackage[all]{hypcap}

\newcommand\msun{\, \rm M_\odot}

\newcommand\kms{\, \rm km\,s^{-1}}

\newcommand\mmbh{{M_{\rm MBH}}}
\newcommand\chimbh{{\chi_{\rm MBH}}}
\newcommand\msbh{{m_{\rm SBH}}}
\newcommand\mnsc{{M_{\rm NSC}}}
\newcommand\rhonsc{{\rho_{\rm NSC}}}
\newcommand\nnsc{{n_{\rm NSC}}}
\newcommand\vkick{{v_{\rm kick}}}
\newcommand\vesc{{v_{\rm esc}}}

\newcommand\be{\begin{equation}}
\newcommand\ee{\end{equation}}

\title[Repeated mergers and ejection of BHs within NSCs]{Repeated mergers and ejection of black holes  within nuclear star clusters}
\author[G. Fragione \& J. Silk]{\parbox{\textwidth}{Giacomo Fragione$^{1,2}$\thanks{E-mail: giacomo.fragione@northwestern.edu}, Joseph Silk$^{3,4,5}$}\\
\ \\
$^1$Department of Physics \& Astronomy, Northwestern University, Evanston, IL 60202, USA\\
$^2$Center for Interdisciplinary Exploration \& Research in Astrophysics (CIERA), Evanston, IL 60202, USA\\
$^3$Institut d’Astrophysique de Paris (UMR7095: CNRS \& UPMC, Sorbonne University), Paris, France\\
$^4$Department of Physics and Astronomy, The Johns Hopkins University, Baltimore, MD, USA\\
$^5$BIPAC, Department of Physics, University of Oxford, Oxford, UK}

\begin{document}

\maketitle

\begin{abstract}
Current stellar evolution models predict a dearth of black holes (BHs) with masses $\gtrsim 50\msun$ and $\lesssim 5\msun$, and intermediate-mass black holes (IMBHs;  $\sim10^2- 10^5\rm M_\odot$) have not yet been detected beyond any reasonable doubt. A natural way to form massive BHs is through repeated mergers, detectable via gravitational wave emission with current LIGO/Virgo or future LISA and ET observations. Nuclear star clusters (NSCs) have masses and densities high enough to retain most of the merger products, which acquire  a recoil kick at the moment of merger. We explore the possibility that IMBHs may be born as a result of repeated mergers in NSCs, and show how their formation pathways depend on the NSC mass and density, and BH spin distribution. We find that  BHs in the pair-instability mass gap can be formed and observed by LIGO/Virgo, and show that the typical mass of the ejected massive BHs is $400$--$500\msun$, with velocities of up to a few thousand $\kms$. Eventually some of these IMBHs  can become the seeds of supermassive BHs,  observed today in the centers of galaxies. In dwarf galaxies, they could potentially solve the  abundance,  core-cusp,  too-big-to-fail,  ultra-faint, and baryon-fraction issues via plausible feedback scenarios.
\end{abstract}


\section{Introduction}
\label{sect:intro}

Black holes (BHs) are commonly subdivided into three different categories, regardless of their spin and charge. Supermassive black holes have masses $\gtrsim 10^5 \msun$ and reside in the centres of galaxies, where they modulate the surrounding gas, star, and compact object distributions \citep{korm2013,alex2017}. Stellar-mass black holes (SBHs) have masses in the range $\sim 5\msun$--$100\msun$ and are the end product of the evolution of massive stars, recently detected by LIGO/Virgo via gravitational wave (GW) emission \citep{abb16,abb17}. Intermediate-mass black holes (IMBHs) have sizes between the previous two categories ($100\msun$--$10^5 \msun$), and there is only circumstantial evidence for their existence \citep{baldass2018,lin18,greene2019}.

Even though LIGO/Virgo has detected several SBHs \citep{LIGO2018b} and dozens of them were known from observations of X-ray binaries \citep{Ozel2010,Farr2011,Corral-Santana2016}, the exact shape of the SBH mass spectrum remains a mystery. Current stellar evolution models predict a dearth of SBHs both with masses $\gtrsim 50\msun$ and $\lesssim 5\msun$ \citep[e.g.,][]{belc2012,fryer2012}. The high-mass gap results from pulsational pair-instabilities affecting the massive progenitors, which can lead to ejection of large amounts of mass whenever the pre-explosion stellar core is $\sim 45\msun - 65\msun$. This would leave a SBH remnant with a maximum mass $\sim 50\msun$ \citep{heger2003,woosley2017,limongi2018,bel2020,mapelli2020}. The high-mass gap extends up to $\sim 125\msun$ \citep{spera2017,renzo2020}, in the nominal IMBH regime. The low-mass gap is related to the explosion mechanism in a core-collapse supernova \citep{belc2012,fryer2012}.

\begin{table*}
\caption{Model parameters: name, seed mass ($M_{\rm seed}$), cluster mass ($\mnsc$), cluster density ($\rhonsc$), spin model, maximum spin ($\chi_{\rm max}$).}
\centering
\begin{tabular}{lccccc}
\hline
Name & $M_{\rm seed}$ ($\msun$) & $\mnsc$ ($10^6 \msun$) & $\rhonsc$ ($10^6 \msun$ pc$^{-3}$) & Spin & $\chi_{\rm max}$\\
\hline\hline
Mod1 & $50$--$200$ & $5$--$500$ & $1$ & uniform & $1.0$\\
Mod1b & $50$--$200$ & $5$--$500$ & $0.1$ & uniform & $1.0$\\
Mod2 & $50$--$200$ & $100$ & $0.1$-$100$ & uniform & $1.0$\\
Mod2b & $50$--$200$ & $10$ & $0.1$-$100$ & uniform & $1.0$\\
Mod3 & $50$--$200$ & $10$ & $1$ & \citet{belc2017} & -\\
\hline
\end{tabular}
\label{tab:models}
\end{table*}

A natural way to form massive SBHs in the pair instability mass gap 
is through repeated mergers of low-mass SBHs. To detect such SBHs through GW emission, the merger remnant has to acquire  a new companion with which to merge, thus requiring a dynamically active environment \citep[e.g.,][]{antoras2016,flr2020}. A fundamental limit for repeated mergers is set by the recoil kick imparted to merger remnants as a result of anisotropic GW emission \citep{lou10,lou11}. Depending on the mass ratio and the spins of the merging objects, the recoil kick could exceed the local escape speed, thus ejecting the system and preventing further mergers \citep{gerosa2019}. For SBHs, a number of studies have shown that massive globular clusters \citep{rodri2019}, nuclear clusters \citep[NSCs;][]{antonini2019}, and AGN disks \citep{mck2020} are the only environments where second-generation mergers can take place, owing to their high escape speed. However, only in the latter two systems is the escape speed  high enough to have possibly more then one consecutive merger.
This  could eventually result in the growth of an IMBH from a SBH seed.

If this mechanism is at work, dwarf galaxies could be naturally populated by IMBHs formed through repeated mergers, which could solve most of the issues related to dwarf galaxies in the $\Lambda$CDM context. The main alternative to mergers as a mechanism for forming IMBH in dwarf galaxies is Bondi-Hoyle-Lyttleton accretion, fed either by infall of dense gas clumps, so-called chaotic accretion, or fuelled by tidal disruption of stars in the NSC \citep{2017MNRAS.467.4180S}. However accretion rates are highly uncertain, being underestimated relative to the
Bondi-Hoyle-Lyttleton rate in some cases by a order of magnitude \citep{2013MNRAS.432.3401G}, are sensitive to numerical resolution \citep{2017MNRAS.467.3475N}, and even  greatly overestimated in some circumstances, for example  where infall-induced outflows dominate, by as much as two orders of magnitude \citep{2019MNRAS.484.1724B} or even more \citep{2020MNRAS.491L..76W}. Hence a dynamical growth mechanism, boosted, as we argue here, by the presence of the NSC, and that involves complementary physics,  merits serious study.

Over the coming years and decades, LIGO/Virgo and forthcoming detectors (e.g., LISA, Einstein Telescope, DECIGO) promise to provide unprecedented constraints on the SBH and IMBH mass distributions \citep{mil2009,ligo_imbh19}. In order to capitalize on the plethora of upcoming observations, it is essential to advance our theoretical understanding of the various pathways through which massive SBHs and IMBHs may form. In this paper, we explore the possibility that IMBHs may be born as a result of repeated mergers of a seed of mass $M_{\rm seed}$ with low-mass SBHs in NSCs. In particular, we quantify how the process depends on the mass and density of the host NSC, and discuss the role played by the initial seed mass and the adopted SBH spin distribution. We also discuss the implications for forming SBHs in the high-mass gap and their possible detection by LIGO/Virgo. Finally, we describe the populations of massive SBHs and IMBHs that are ejected during this process.

The paper is organized as follows. In Section \ref{sect:method}, we describe the method we use to dissect the origin of repeated mergers and ejections of black holes in NSCs, while, in Section \ref{sect:results}, we describe the results of our investigation. Finally, we discuss the implications of our findings for gravitational-wave astrophysics, galactic astrophysics, and cosmology and draw our conclusions in Section \ref{sect:conc}.

\section{Method}
\label{sect:method}

We consider the formation of a massive BH (MBH)\footnote{We refer to a MBH either as a SBH in the pair-instability mass gap or as an IMBH or SMBH in the range above $\sim 100\rm M_\odot$.} of mass $\mmbh$ that repeatedly merges with SBHs of mass $5 \msun \le m_2 \le 50 \msun$, starting from a seed of mass $M_{\rm seed}$ ($\mmbh(T=0)=M_{\rm seed}$), in the core of a NSC. To explore the role of the seed mass, we specify  it to span four different initial values, $50\msun$--$100\msun$--$150\msun$--$200\msun$. The mass of $m_2$ is drawn assuming that the pairing probability for its SBH components scales as $\mathcal{P}\propto (\mmbh+m_2)^4$, as is  appropriate for binaries formed via three-body processes \citep{omk16}.

The characteristics of the host NSCs are essentially determined by their mass $\mnsc$ and half-mass mass density $\rhonsc$ \citep{antoras2016}. The escape velocity as a function of cluster mass and density can be straightforwardly computed \citep{georg2009}
\begin{equation}
\vesc=40\kms \left(\frac{\mnsc}{10^5\msun}\right)^{1/3} \left(\frac{\rhonsc}{10^5\msun\mathrm{pc}^{-3}}\right)^{1/6}\,,
\label{eqn:mvesc}
\end{equation}
with the cluster velocity dispersion being $\sigma=\vesc/(2\sqrt{3})$.

In the NSC environment, SBHs segregate to the cluster core where stellar binaries form and shrink through three-body encounters. These binaries will eventually have semi-major axis and eccentricities such that they will merge through emission of GW radiation. In our study, we consider the merger as a three-step process.

Firstly, we assume that, after binaries are formed, they will dominate the dynamics inside the cluster core. Interactions come in two flavours, encounters between three single objects and encounters between a single and a binary. The typical time-scale for the former is \citep[e.g.,][]{lee1995}
\begin{eqnarray}
t_{\rm 3bb}&=&125\ \mathrm{Myr}\left(\frac{10^6\ \mathrm{pc}^{-3}}{\nnsc}\right)^2\times \nonumber\\
&\times& \left(\xi^{-1} \frac{\sigma}{30\kms}\right)^9 \left(\frac{20\msun}{\mmbh}\right)^5\,,
\end{eqnarray}
where $\nnsc$ is the number density of BHs near the center and
\begin{equation}
\xi^{-1}=\frac{\langle \msbh\rangle\sigma_{\rm BH}^2}{\langle m_{\rm *}\rangle \sigma^2}\,,
\end{equation}
is the deviation from energy equipartition between SBHs and stars in the NSC. We fix $\xi^{-2}=0.2$ \citep{mors2015}. Encounters between a single and a binary occur on a time-scale\footnote{Assuming a SBH typical mass of $10\msun$ and a binary fraction $f_b=0.01$ \citep{mors2015}. If the cluster core is dominated by stellar binaries, binary BH formation is first mediated by the interaction of BHs with binary stars \citep{antoras2016}.} \citep{mill2009}
\begin{eqnarray}
t_1&=&300\ \mathrm{Myr}\ \xi^{-1}\left(\frac{10^6\ \mathrm{pc}^{-3}}{\nnsc}\right)\left(\frac{\sigma}{30\kms}\right)\times\nonumber\\
&\times& \left(\frac{30\msun}{10\msun+\mmbh+m_2}\right)\left(\frac{1\ \mathrm{AU}}{a_h}\right)\,,
\end{eqnarray}
where the hardening semi-major axis \citep{quin1996}
\begin{equation}
a_h=1\ \mathrm{AU}\left(\frac{m_2}{4\msun}\right)\left(\frac{\sigma}{30\kms}\right)^{-2}\,.
\end{equation}

\begin{figure} 
\centering
\includegraphics[scale=0.55]{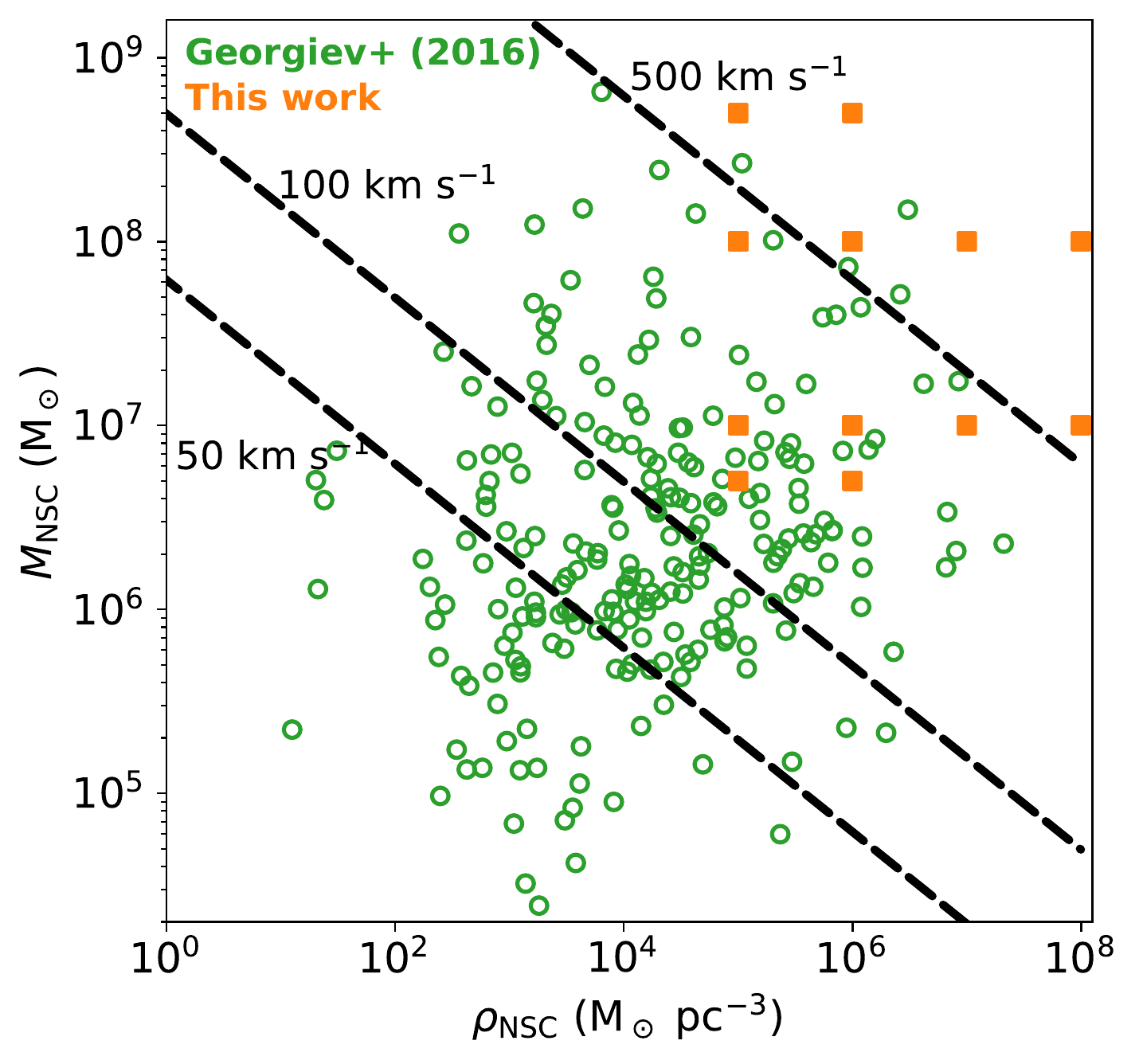}
\caption{Observed mass and half-mass density of NSCs from \citet{georg2016}. Orange squares represent the models investigated in this work. Black-dashed lines show different escape speeds, computed using Eq.~\ref{eqn:mvesc}.}
\label{fig:datavssim}
\end{figure}

Secondly, the binary shrinks at a constant rate \citep{quin1996}, eventually to the regime where GWs takes over. However, during one of the interactions that makes the binary shrink, the binary itself can receive a dynamical kick such that it is ejected. This happens whenever \citep{antoras2016}
\begin{equation}
a_{\rm ej}>a_{\rm GW}\,,
\label{eqn:ejok}
\end{equation}
where
\begin{eqnarray}
a_{\rm ej}&=&0.07\ \mathrm{AU}\left(\frac{100\ \mathrm{M}_\odot^2}{(\mmbh+m_2)(10\msun+\mmbh+m_2)}\right)\times\nonumber\\
&\times&\left(\frac{\mu}{1\msun}\right)\left(\frac{50\kms}{\vesc}\right)^2\,,
\end{eqnarray}
where $\mu$ is the reduced mass of $\mmbh$-$m_2$, and
\begin{eqnarray}
a_{\rm GW}&=&0.05\ \mathrm{AU}\left(\frac{\mmbh+m_2}{20\msun}\right)^{3/5}\left(\frac{10^6\msun\ \mathrm{pc}^{-3}}{\rhonsc}\right)^{1/5}\times \nonumber\\
&\times& \left(\frac{\sigma}{30\kms}\right)^{1/5}\left(\frac{q}{(1+q)^2}\right)^{1/5}\,,
\end{eqnarray}
where $q=m_2/\mmbh$. The binary will shrink until reaching $\max(a_{\rm ej},a_{\rm GW})$ over a time-scale \citep{millh2002,gult2006}
\begin{eqnarray}
t_2&=&200\ \mathrm{Myr}\ \xi^{-1}\left(\frac{10^6\ \mathrm{pc}^{-3}}{\nnsc}\right)\left(\frac{\sigma}{30\kms}\right)\times\nonumber\\
&\times &\left(\frac{0.05\ \mathrm{AU}}{\max(a_{\rm ej},a_{\rm GW})}\right)\left(\frac{10\msun}{\mmbh+m_2}\right)^2
\end{eqnarray}
at a typical semi-major axis. If Eq.~\ref{eqn:ejok} is verified, the binary is ejected from the cluster and the eventual merger takes place outside of the NSC, quenching further growth. We sample all the relevant time-scales from a Poisson distribution, that is $\exp(-t/\tau)$, with $\tau=t_{\rm 3bb}$, $t_1$, $t_2$.

Thirdly, the binary merges over a time-scale \citep{peters64}
\begin{eqnarray}
T_{\rm GW}&=&2000\ \mathrm{Myr}\left(\frac{10^3\msun}{\mmbh m_2(\mmbh+m_2)}\right)\times \nonumber\\
&\times& \left(\frac{\max(a_{\rm ej},a_{\rm GW})}{0.05\ \mathrm{AU}}\right)^4 (1-e^2)^{3/2}
\end{eqnarray}
where $e$ is the eccentricity, that we sample from a thermal distribution, achieved through many dynamical encounters \citep{jeans1919,Heggie1975}. If $a_{\rm ej}>a_{\rm GW}$, the merger will take place outside of the cluster, otherwise within the cluster. If it happens outside of the cluster, the MBH is considered ejected and further growth is prevented.

As a result of the anisotropic emission of GWs at merger, a recoil kick is imparted to the merger remnant \citep{lou12}, which can eject it from the host NSC. The recoil kick depends on the asymmetric mass ratio $\eta=q/(1+q)^2$ and on the magnitude of the reduced spins, $|\mathbf{{\chi_{\rm MBH}}}|$ and $|\mathbf{{\chi_2}}|$.
We model the recoil kick as \citep{lou10}
\begin{equation}
\textbf{v}_{\mathrm{kick}}=v_m \hat{e}_{\perp,1}+v_{\perp}(\cos \xi \hat{e}_{\perp,1}+\sin \xi \hat{e}_{\perp,2})+v_{\parallel} \hat{e}_{\parallel}\,,
\label{eqn:vkick}
\end{equation}
where
\begin{eqnarray}
v_m&=&A\eta^2\sqrt{1-4\eta}(1+B\eta)\\
v_{\perp}&=&\frac{H\eta^2}{1+q}(\chi_{2,\parallel}-q\chi_{1,\parallel})\\
v_{\parallel}&=&\frac{16\eta^2}{1+q}[V_{1,1}+V_A \tilde{S}_{\parallel}+V_B \tilde{S}^2_{\parallel}+V_C \tilde{S}_{\parallel}^3]\times \nonumber\\
&\times & |\mathbf{\chi}_{2,\perp}-q\mathbf{\chi}_{1,\perp}| \cos(\phi_{\Delta}-\phi_{1})\,.
\end{eqnarray}
The $\perp$ and $\parallel$ refer to the direction perpendicular and parallel to the orbital angular momentum, respectively, while $\hat{e}_{\perp}$ and $\hat{e}_{\parallel}$ are orthogonal unit vectors in the orbital plane. We have also defined the vector
\begin{equation}
\tilde{\mathbf{S}}=2\frac{\mathbf{\chi}_{2,\perp}+q^2\mathbf{\chi}_{1,\perp}}{(1+q)^2}\,,
\end{equation}
$\phi_{1}$ as the phase angle of the binary, and $\phi_{\Delta}$ as the angle between the in-plane component of the vector
\begin{equation}
\mathbf{\Delta}=M^2\frac{\mathbf{\chi}_{2}-q\mathbf{\chi}_{1}}{1+q}
\end{equation}
and the infall direction at merger. Finally, we adopt $A=1.2\times 10^4$ km s$^{-1}$, $H=6.9\times 10^3$ km s$^{-1}$, $B=-0.93$, $\xi=145^{\circ}$ \citep{gon07,lou08}, and $V_{1,1}=3678$ km s$^{-1}$, $V_A=2481$ km s$^{-1}$, $V_B=1793$ km s$^{-1}$, $V_C=1507$ km s$^{-1}$ \citep{lou12}. We adjust the final total spin of the merger product and its mass as in \citet{rezzolla2008}. 

As a result of the kick, the remnant could be ejected from the host NSC. This happens whenever $v_{\rm kick}>v_{\rm esc}$. On the other hand, if $v_{\rm kick}<v_{\rm esc}$, $\mmbh$ will be retained within the NSC and will get to a distance\footnote{\citet{antonini2019} use $R_{\rm ej}=\sqrt{v_{\rm esc}^4/(v_{\rm esc}^2-v_{\rm kick}^2)^2-1}$, which is valid for a Plummer model and also provides a good approximation for moderately concentrated King models.}
\begin{equation}
R_{\rm ej}=r_h\left(\frac{\vkick}{\vesc}\right)^2\,,
\end{equation}
where $r_h$ is the NSC half-mass radius. When not in the core, the growing MBH do not form efficiently binaries with the other SBHs. However, after a dynamical friction time-scale \citep{binntrem87}
\begin{equation}
\tau_{\rm df}\approx 1\ \mathrm{Myr} \left(\frac{100\msun}{\mmbh}\right)\left(\frac{\mnsc}{10^5\msun}\right)\left(\frac{10^5\msun\ \mathrm{pc}^{-3}}{\rhonsc}\right)^{1/2}\,,
\end{equation}
the merged remnant returns to the cluster core and can start interacting again with other SBHs, eventually growing further in mass.

\begin{figure} 
\centering
\includegraphics[scale=0.575]{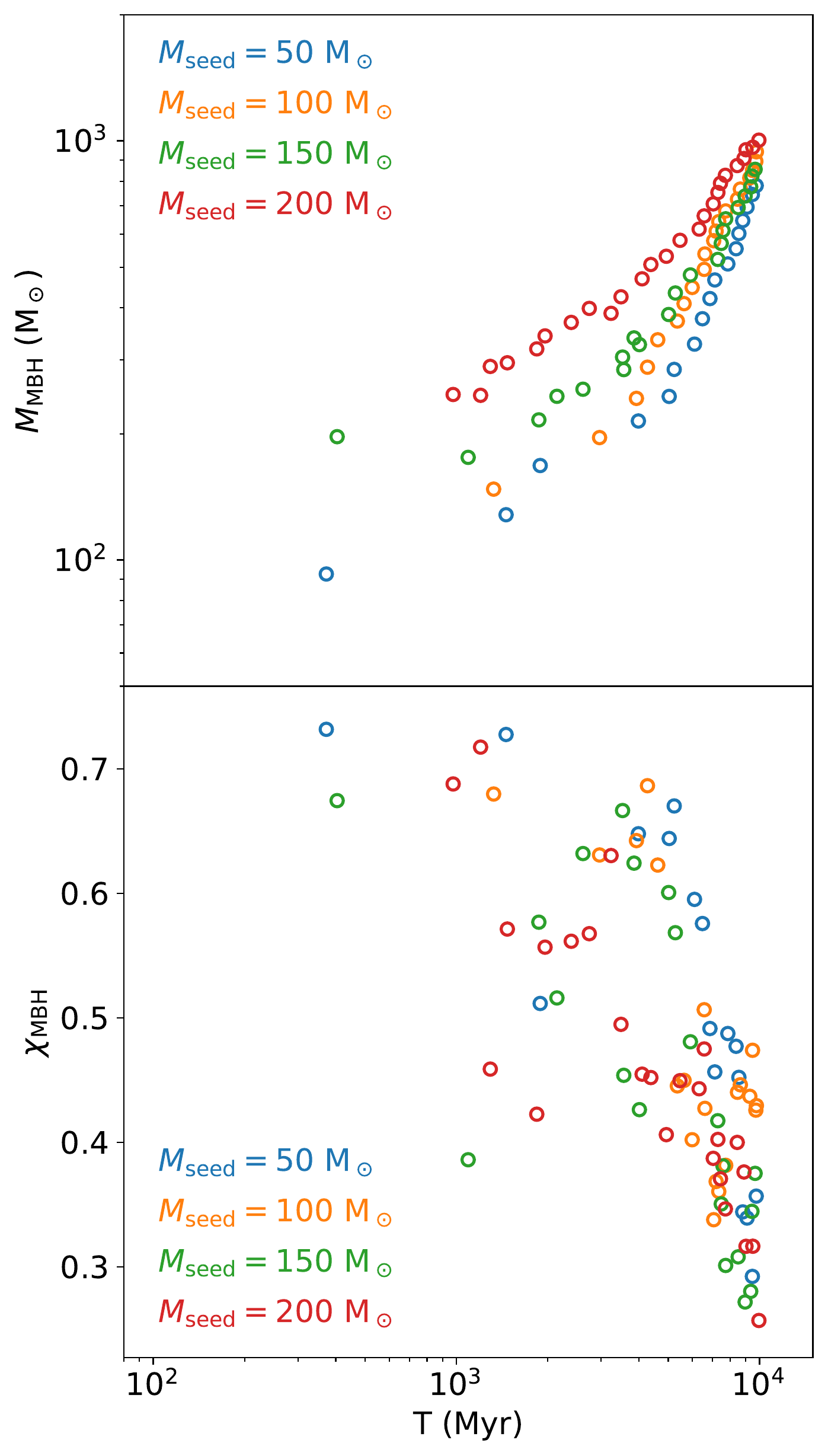}
\caption{Example of growth of a $\sim 1000\msun$ IMBH ($\mmbh$) starting from a seed mass $M_{\rm seed}=50$--$200\msun$. Different seeds correspond to different velocities in the growth, while the spin tends to $\lesssim 0.4$. In this example, the mass and density of the NSC are $\mnsc=10^8\msun$ and $\rhonsc=10^6\msun$ pc$^{-3}$, respectively.}
\label{fig:examp}
\end{figure}

\begin{figure*} 
\centering
\includegraphics[scale=0.46]{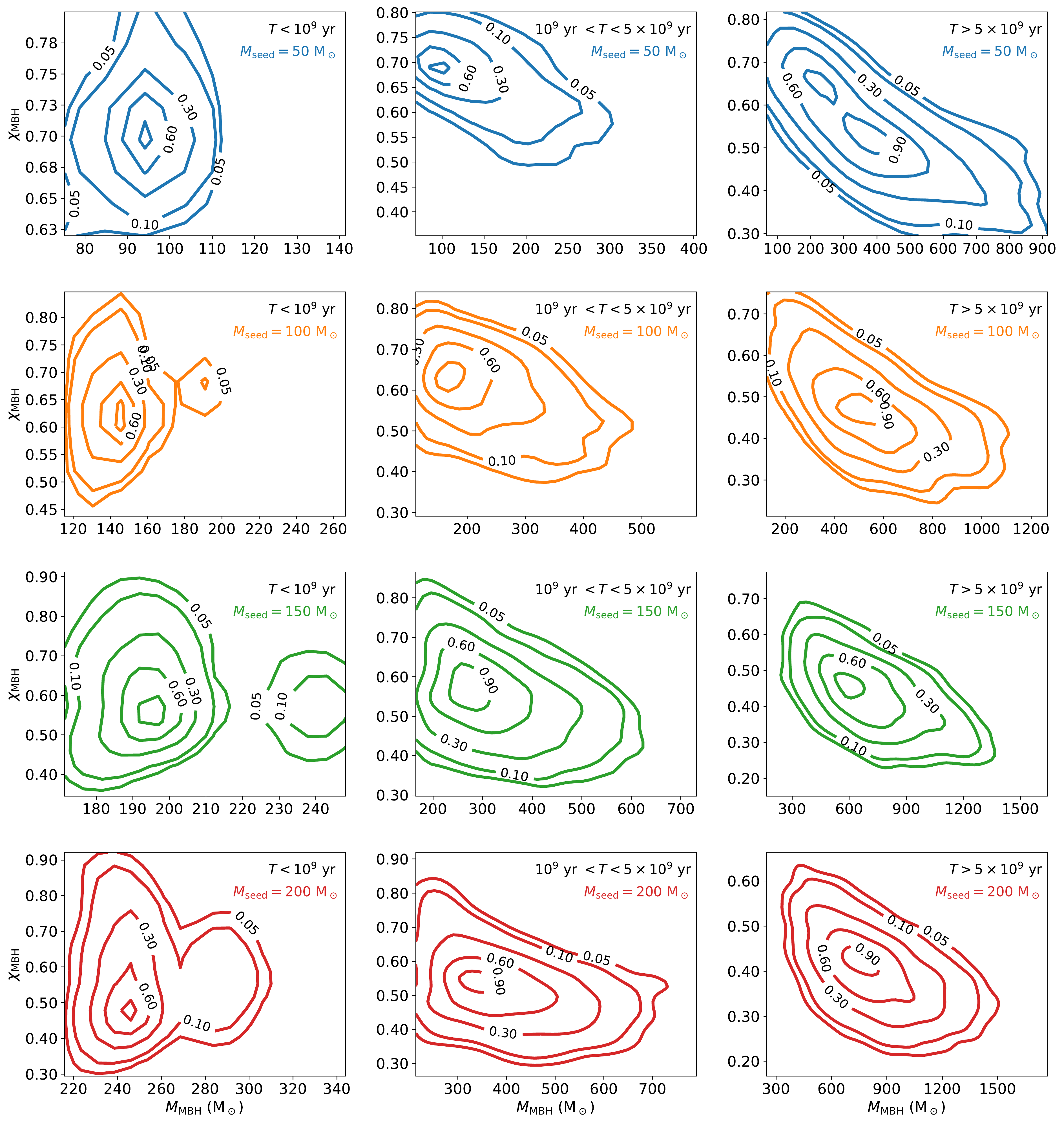}
\caption{Distribution of the mass and spin of MBHs retained in a NSC of mass $\mnsc=10^8\msun$ and density $\rhonsc=10^6\msun$ pc$^{-3}$ at different epochs. Left panel: $T<10^9$ yr; central panel: $10^9$ yr $<T<5\times 10^9$ yr; right panel: $T>5\times 10^9$ yr.}
\label{fig:retained}
\end{figure*}

\begin{figure*} 
\centering
\includegraphics[scale=0.75]{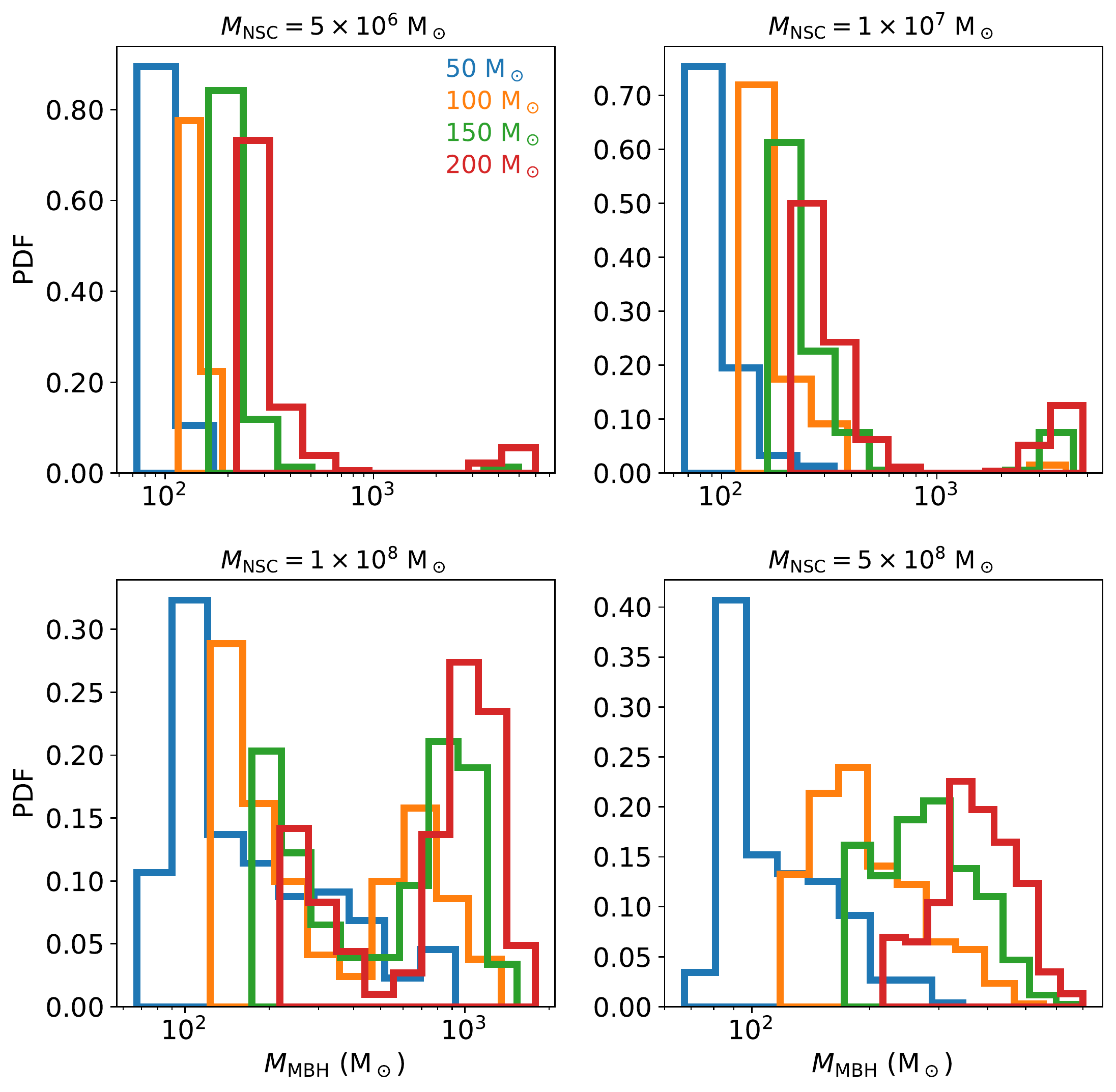}
\caption{Probability distribution functions of the final mass of MBHs obtained for different NSC masses: top-left $M_{\rm NSC}=5\times 10^6\msun$; top-right $M_{\rm NSC}=1\times 10^7\msun$; bottom-left $M_{\rm NSC}=1\times 10^8\msun$; bottom-right $M_{\rm NSC}=5\times 10^8\msun$. The NSC density is $\rho_{\rm NSC}=10^6\msun$ pc$^{-3}$. Different colors represent different seed masses.}
\label{fig:mod1}
\end{figure*}

\begin{figure*} 
\centering
\includegraphics[scale=0.75]{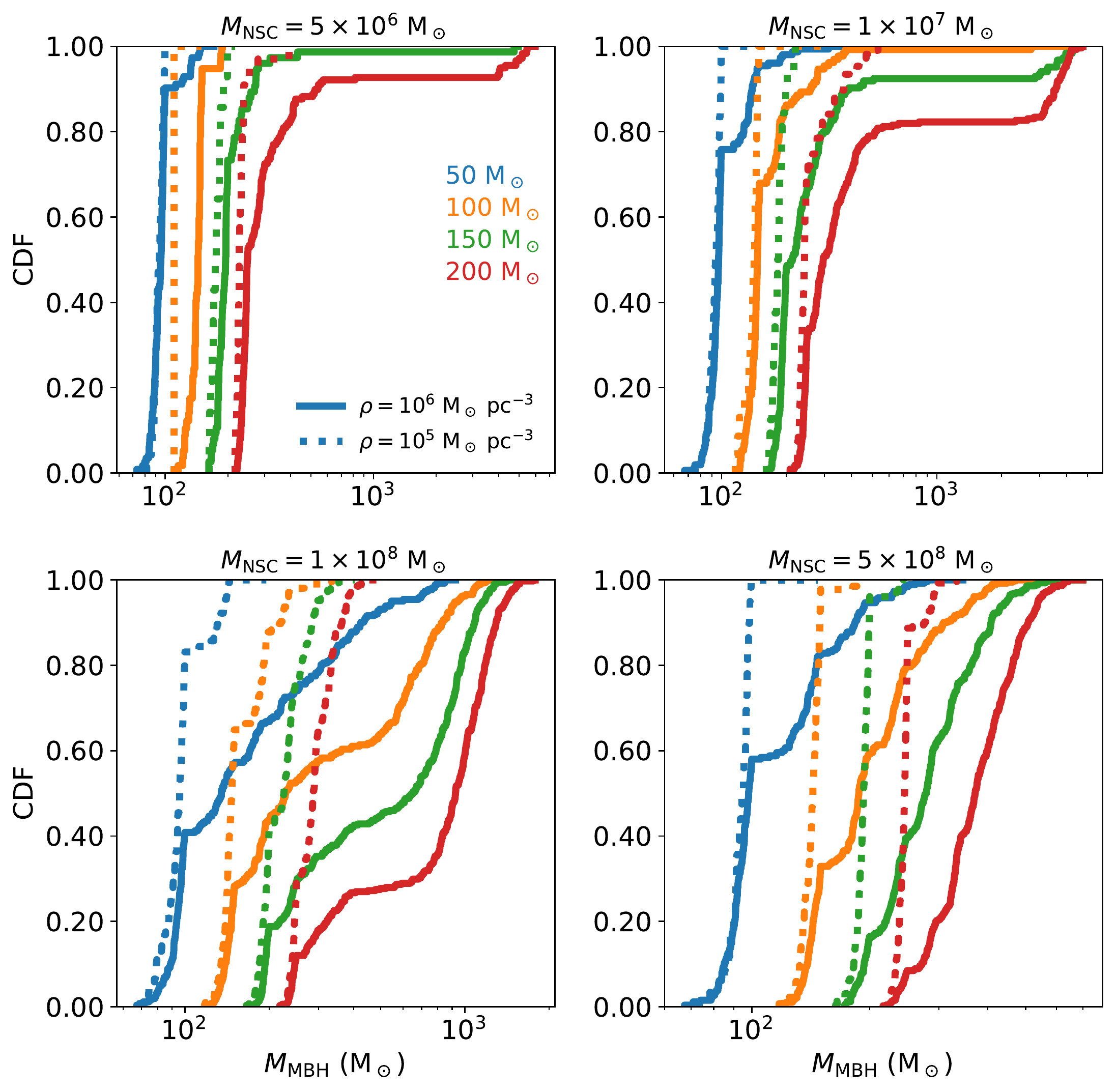}
\caption{Comparison of the cumulative distribution function of the final mass of MBHs for different NSC masses: top-left $M_{\rm NSC}=5\times 10^6\msun$; top-right $M_{\rm NSC}=1\times 10^7\msun$; bottom-left $M_{\rm NSC}=1\times 10^8\msun$; bottom-right $M_{\rm NSC}=5\times 10^8\msun$. The NSC density is $\rho_{\rm NSC}=10^6\msun$ pc$^{-3}$ (solid line) and $\rho_{\rm NSC}=10^5\msun$ pc$^{-3}$ (dashed line). Different colors represent different seed masses.}
\label{fig:compmod1}
\end{figure*}

\begin{figure*} 
\centering
\includegraphics[scale=0.75]{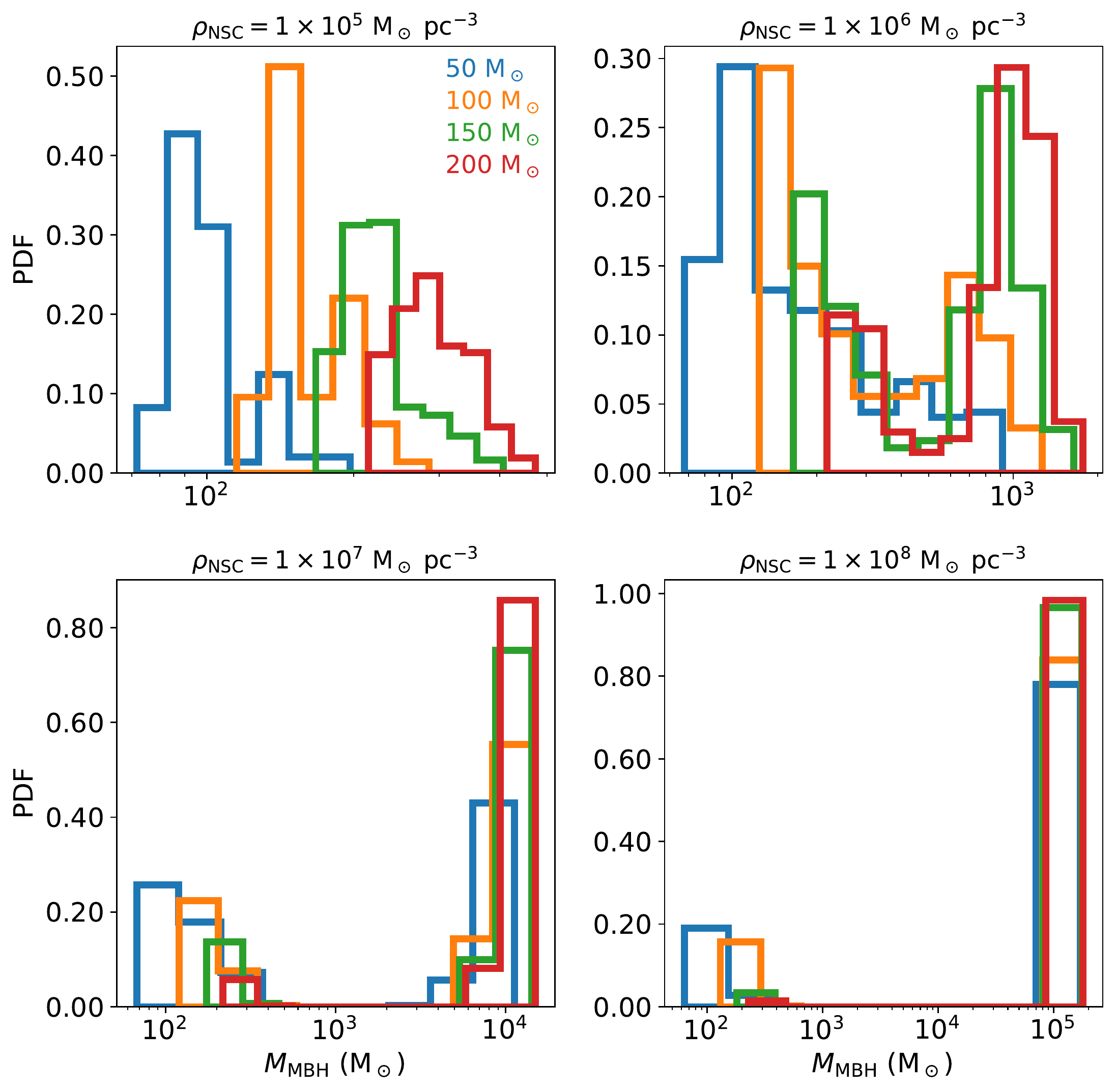}
\caption{Probability distribution functions of the final mass of MBHs obtained for different NSC densities: top-left $\rho_{\rm NSC}=1\times 10^5\msun$ pc$^{-3}$; top-right $\rho_{\rm NSC}=1\times 10^6\msun$ pc$^{-3}$; bottom-left $\rho_{\rm NSC}=1\times 10^7\msun$ pc$^{-3}$; bottom-right $\rho_{\rm NSC}=1\times 10^8\msun$. The NSC density is $M_{\rm NSC}=10^8\msun$ pc$^{-3}$. Different colors represent different seed masses.}
\label{fig:mod2}
\end{figure*}

\begin{figure*} 
\centering
\includegraphics[scale=0.675]{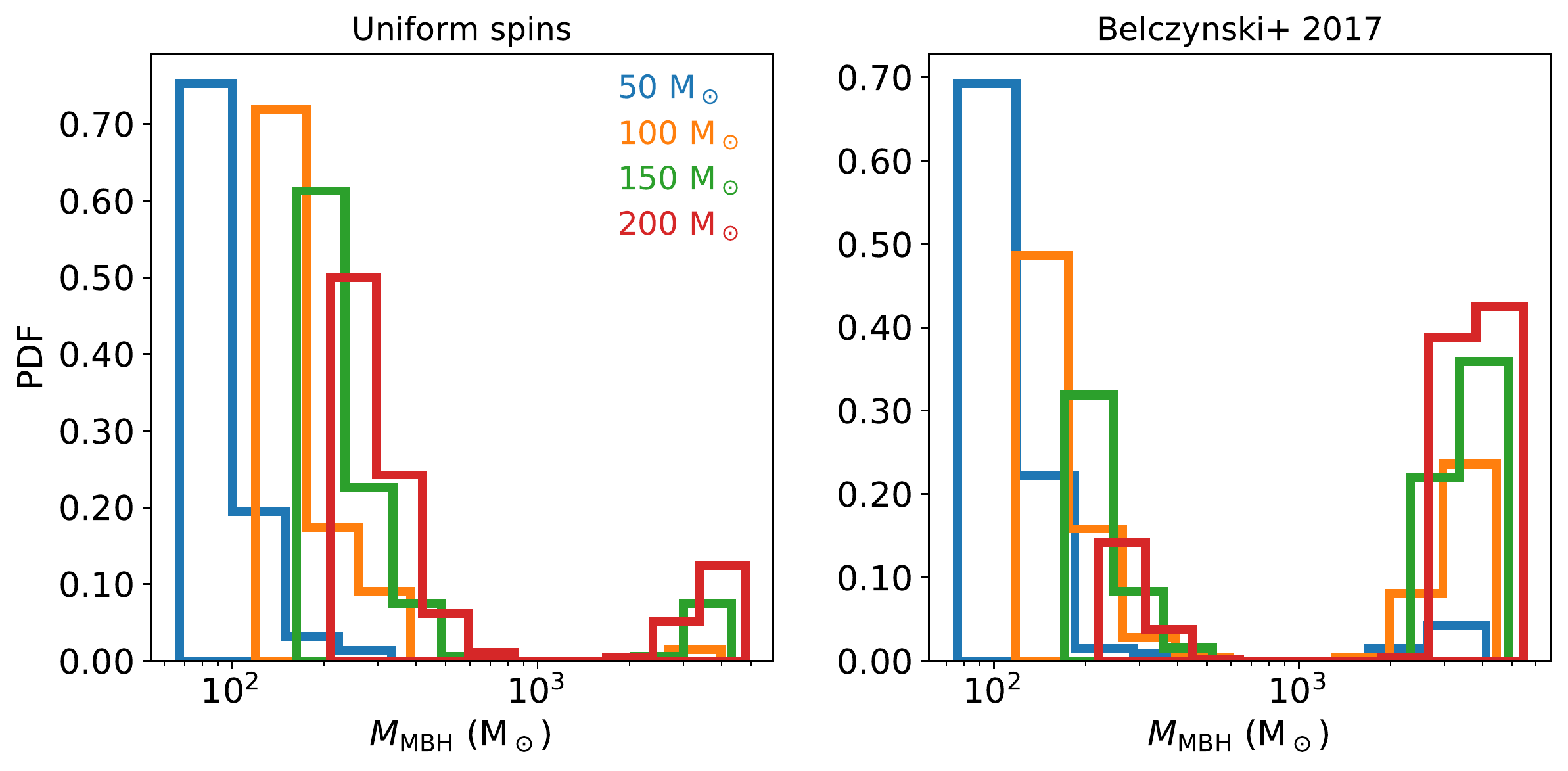}
\caption{Probability distribution functions of the final mass of MBHs obtained for $M_{\rm NSC}=1\times 10^7\msun$ pc$^{-3}$ and $\rho_{\rm NSC}=1\times 10^6\msun$ pc$^{-3}$ for a uniform distribution of SBH spins (left) and a \citet{belc2017} distribution of spins (right). Different colors represent different seed masses.}
\label{fig:mod4}
\end{figure*}

\section{Results}
\label{sect:results}

We summarize in Table~\ref{tab:models} the models we investigate in our simulations. To explore the role of the seed mass ($\mmbh(T=0)=M_{\rm seed}$), we use four different mass values ($50\msun$, $100\msun$, $150\msun$, $200\msun$) for each model. We consider different masses and densities of NSCs, and different spin models. NSC masses and densities are chosen such that their escape speed (see Eq.~\ref{eqn:mvesc}) is high enough to efficiently retain the massive seed undergoing repeated mergers \citep{antonini2019}. Spin orientations are assumed to be isotropic, as appropriate for merging binaries assembled dynamically. We stop each simulation either if the maximum integration time of $T=10^{10}$ yr is reached or if the MBH is ejected via dynamical or recoil kick. Each model is the average over $1000$ realizations.

We show in Figure~\ref{fig:datavssim} the total mass and density of NSCs in the sample of \citet{georg2016}. This sample of NSCs comprises the systems in spheroid-dominated galaxies \citep{cote2006,turner2012} and disc-dominated galaxies \citep{georg2009,geor2014}. We also overplot the mass and the density of the NSCs  considered in our models. For reference, we show the lines of the relative escape speed ($50\kms$, $100\kms$, $500\kms$), computed using Eq.~\ref{eqn:mvesc}. The recoil kick velocity imparted to a growing MBH depends on the asymmetric mass ratio $\eta$ and on the intrinsic spins of the progenitors \citep[e.g.,][]{fragk18}. While for lower mass ratios, the recoil kick is smaller, as the spin of the merging objects plays a crucial role (see Eq.~\ref{eqn:vkick}). Consequently, even for modest mass ratios, $\vkick$ can be as large as a few hundred $\kms$ \citep[e.g.,][]{fragleiginkoc18}, enough to eject any merger remnant from a NSC. Only NSCs with local escape speeds $\vesc\gtrsim 300\kms$--$400\kms$ may retain the merger products, which can later merge again and form a MBH \citep{antonini2019}. However, NSCs densities could be likely higher at high redshifts which would favor the retention of MBHs and their growth.

\subsection{Growth of intermediate-mass black holes}
\label{subsect:growth}
The pathway to form a MBH is quite general, regardless of the initial seed mass. As $\mmbh$ grows, $\eta$ gets smaller, thus $\vkick$ decreases. However, intrinsic spins can still play an important role in determining the fate of the merger remnant. High spins favor large recoil kicks, which can be large enough to eject the MBH even from massive NSCs. Therefore, the probability of retaining any seed that is going to grow to large masses increases for small intrinsic spins.

For equal mass mergers, the final spin is peaked at $\sim 0.7$ independently of the initial spins, due to the orbital angular momentum of the merging objects \citep{hof16}. In general, the angular momentum parameter ($\chi_{\rm merg}$) of a BH that is the merger product of $\mmbh$ and $m_2$ is the sum of three contributions \citep{buo08}
\begin{equation}
\chi_{\rm merg}=\frac{L_{\mathrm{orb}}(\mu,r_{\mathrm{ISCO}},\chi_{\rm merg})}{M^3}+\frac{M^3_{\mathrm{MBH}}\chi_{\mathrm{MBH}}}{M^3}+\frac{M^3_{\mathrm{2}}\chi_{\mathrm{2}}}{M^3}\,,
\label{eqn:finspin}
\end{equation}
where $M=\mmbh+m_2$ is the binary total mass, $\chi_{\mathrm{MBH}}$ and $\chi_{\mathrm{2}}$ are the reduced spins of $\mmbh$ and $m_2$, respectively, and $L_{\mathrm{orb}}(\mu,r_{\mathrm{ISCO}},\chi_{\rm merg})$ is the orbital angular momentum of a particle of mass $\mu$ (reduced mass of $\mmbh$ and $m_2$) at the ISCO of a Kerr black hole of spin parameter $\chi_{\rm merg}$. If $M_{\mathrm{MBH}}\gg m_{\mathrm{2}}$, the spin of the merger product is essentially dominated by the contribution of the growing MBH (with some contribution from the angular momentum). Since the probability of retaining any seed increases for small intrinsic spins, the growing MBH is expected to have a low $\chimbh$. Once the condition $M_{\mathrm{MBH}}\gg m_{\mathrm{2}}$ is achieved, the spin of the MBH essentially freezes out and $\vkick$ will constantly stay below the escape speed, allowing the MBH to efficiently continue to grow in mass.

We show an example of this in Figure~\ref{fig:examp}, where we plot the mass and the spin of a MBH growing up to $\sim 1000\msun$ IMBH ($\mmbh$), starting from a seed mass $M_{\rm seed}=50$--$200\msun$. In this example, the mass and density of the cluster are $\mnsc=10^8\msun$ and $\rhonsc=10^6\msun$ pc$^{-3}$, respectively. Different seeds correspond to different velocities in the MBH growth. However, the path of $\chimbh$ is quite similar among the four different cases, tending to a low value in order that $\vkick$ becomes small enough to retain the growing MBH. 

For the same motivations discussed above, we also expect a correlation between the elapsed time since the seed started growing and the MBH intrinsic spin. We illustrate this concept in Figure~\ref{fig:retained}, where we plot the distribution of the mass and spin of MBHs retained in a NSC of mass $\mnsc=10^8\msun$ and density $\rhonsc=10^6\msun$ pc$^{-3}$ at different epochs. We subdivide the MBH into three time bins: $T<10^9$ yr (left panel), $10^9$ yr $<T<5\times 10^9$ yr (central panel), and $T>5\times 10^9$ yr (right panel). The initial spin of $M_{\rm seed}$ and the the spins of the spins of the SBH that merge with the growing MBH are sampled uniformly in the range $[0,1)$. At early times, $\chimbh$ still spans very high values. As time passes and the MBH is assembled, the MBH spin tends to lower values. At late times, the bulk of the population has typically $\chimbh\lesssim 0.4$.

Figure~\ref{fig:mod1} shows the probability distribution functions of the final mass of MBHs obtained from different initial seed masses for NSCs of mass $M_{\rm NSC}=5\times 10^6\msun$ (top-left), $M_{\rm NSC}=1\times 10^7\msun$ (top-right), $M_{\rm NSC}=1\times 10^8\msun$ (bottom-left), and $M_{\rm NSC}=5\times 10^8\msun$ (bottom-right). The NSC density is $10^6\msun$ pc$^{-3}$. We find that only $\sim 0.4\%$--$9.8\%$ of the initial systems end up assembling an MBH of $\sim 4\times 10^3\msun$ for $M_{\rm NSC}\le 1\times 10^7\msun$. The majority of the seeds would have grown up to a few hundred solar masses by $10^{10}$ yr. For $M_{\rm NSC}= 1\times 10^8\msun$, the distributions tend to become double-peaked. We find that a larger fraction (up to $\sim 52\%$ of the initial systems for $M_{\rm seed}=200\msun$) of MBHs end up having a mass of $\sim 10^3\msun$, smaller than in the previous two cases. The distributions turn back into a single-peaked shape for $M_{\rm NSC}=5\times 10^8\msun$, with a peak that depends on the initial seed mass: the larger $M_{\rm seed}$, the larger is the peak MBH mass.

The previous distributions can be understood in the light of the equations that govern MBH dynamics. MBH growth is a delicate balance between the ability of a NSC to retain the merger remnants when a recoil kick is imparted and the rapidity at which the retained MBHs can form and merge in new binaries. Generally, more massive and denser NSCs have larger escape velocities, as evident from Eq.~\ref{eqn:mvesc}, and can  more efficiently retain the recoiled MBHs. The time-scales for the formation of a binary can be recast in terms of the NSC mass and density
\begin{equation}
t_{\rm 3bb}\propto \frac{M^{3}}{\rho^{1/2}}
\end{equation}
and
\begin{equation}
t_1\propto \frac{M}{\rho^{1/2}}\,,
\end{equation}
for encounters of three singles and binary-single encounters, respectively. The time-scale for shrinking the orbit of the formed binaries can be rewritten as
\begin{equation}
t_2\propto \frac{M^{1/3}}{\rho^{5/6}}\,.
\end{equation}
Therefore, the formation time-scale, as well as the typical time necessary to reduce the orbit of the binary to the GW regime, are larger for larger NSC masses and lower NSC densities. For equal densities, more massive NSCs will form and merge binaries on typically longer time-scales, while, for equal masses, denser NSCs will form and merge binaries on shorter time-scales. As a result, a smaller number of seeds can grow to high masses for smaller NSC masses, their escape speed being smaller. Nevertheless, these seeds will typically reach larger masses since the time-scale to form and merge binaries is smaller than for massive clusters. For massive NSCs, the situation is the other way round, with a larger fraction of seeds able to grow, but to smaller masses since their typical evolutionary time-scales to assemble and merge binaries are longer.

We compare in Figure~\ref{fig:compmod1} the cumulative distribution functions of the final masses of MBHs obtained from different initial seed masses for NSCs of mass $M_{\rm NSC}=5\times 10^6\msun$ (top-left), $M_{\rm NSC}=1\times 10^7\msun$ (top-right), $M_{\rm NSC}=1\times 10^8\msun$ (bottom-left), and $M_{\rm NSC}=5\times 10^8\msun$ (bottom-right). The NSC density is $10^6\msun$ pc$^{-3}$ (solid line) or $10^5\msun$ pc$^{-3}$ (dashed line). In the latter case, the escape speed is $\sim 1.5$ smaller (see Eq.~\ref{eqn:mvesc}) and more MBHs are ejected. As a consequence, MBHs can grow less efficiently to large masses and on typically longer time-scales.

Figure~\ref{fig:mod2} shows the probability distribution functions of the final masses of MBHs obtained for different seed masses and NSC densities of $\rho_{\rm NSC}=1\times 10^5\msun$ pc$^{-3}$ (top-left), $\rho_{\rm NSC}=1\times 10^6\msun$ pc$^{-3}$ (top-right), $\rho_{\rm NSC}=1\times 10^7\msun$ pc$^{-3}$ (bottom-left), $\rho_{\rm NSC}=1\times 10^8\msun$ (bottom-right). In these models, the NSC density is fixed to $M_{\rm NSC}=10^8\msun$ pc$^{-3}$. The shapes of these distributions follow quite nicely the general trends discussed above. For NSC densities $\gtrsim 1\times 10^7\msun$ pc$^{-3}$, the escape speed is $\gtrsim 600\kms$, typically larger than the recoil kick. Moreover, the high NSC densities shorten  the typical time-scale for the formation and merger of binaries. As a result, most of the seeds can significantly grow in mass, up to $\sim 10^4$--$10^5\msun$.

To understand the importance of the spin distributions, we run an additional model where the intrinsic SBH spins are sampled from \citep{belc2017}
\begin{equation}
\chi_{\rm BH}=\frac{p_1-p_2}{2}\tanh\left(p_3-\frac{M_{\rm BH}}{\msun}\right)+\frac{p_1+p_2}{2}\,,
\label{eqn:bhspin}
\end{equation}
where $p_1=0.86\pm 0.06$, $p_2=0.13\pm 0.13$, and $p_3=29.5\pm 8.5$. Spins are generated by drawing random samples uniformly in the region in between the two curves given by the upper and lower limits of the parameters \citep{gerosa2018}. This distribution assigns on average large spins to low-mass SBHs and small spins to high-mass SBHs. Therefore, the fraction of MBH seeds that can efficiently build up mass is expected to be larger than the case in which the SBH spins are sampled uniformly. Figure~\ref{fig:mod4} reports a comparison of the final masses of MBHs obtained for $M_{\rm NSC}=1\times 10^7\msun$ pc$^{-3}$ and $\rho_{\rm NSC}=1\times 10^6\msun$ pc$^{-3}$ for a uniform distribution of SBH spins and a \citet{belc2017} distribution of spins. We find that in the latter case the peak at $\sim 4\times 10^3\msun$ is a factor of $\sim 4$ more pronounced compared to the case when the spins are assumed to be uniformly distributed. This immediately comes from Eq.~\ref{eqn:vkick}, where the recoil kicks are smaller for low spins. Therefore, if the majority of SBHs were born with low spins \citep{fuller2019}, even small seeds can grow up to several thousands of solar masses in a Hubble time.

\subsection{Mergers of black holes in the mass gap}
\label{subsect:ligov}

\begin{figure*} 
\centering
\includegraphics[scale=0.7]{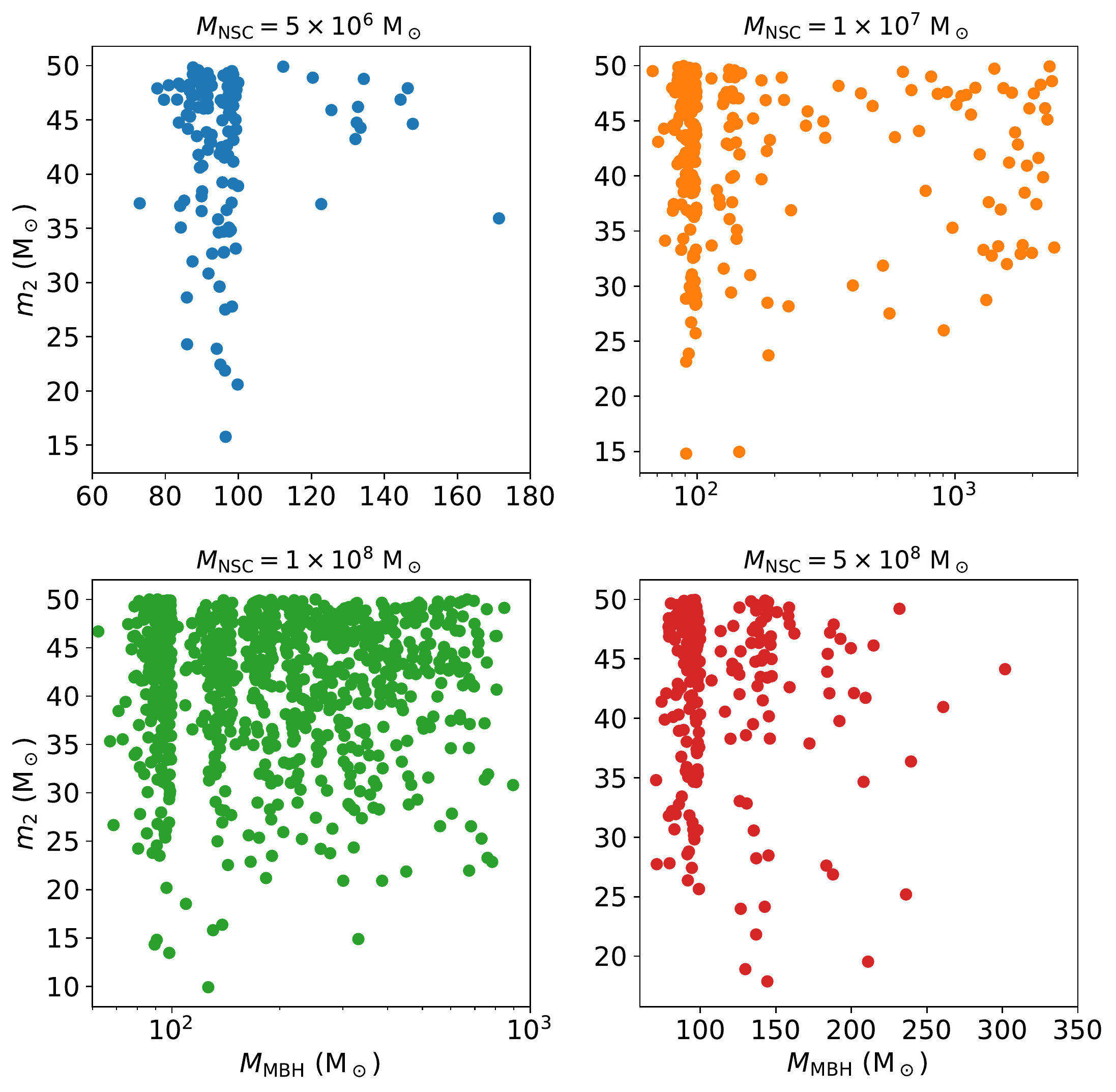}
\caption{Distribution of mergers of MBHs for different NSC masses: top-left $M_{\rm NSC}=5\times 10^6\msun$; top-right $M_{\rm NSC}=1\times 10^7\msun$; bottom-left $M_{\rm NSC}=1\times 10^8\msun$; bottom-right $M_{\rm NSC}=5\times 10^8\msun$. The NSC density is $\rho_{\rm NSC}=10^6\msun$ pc$^{-3}$. The seed mass is $M_{\rm seed}=50\msun$, consistent with the lower boundary of the mass-gap.}
\label{fig:ligov}
\end{figure*}

Current stellar evolution models predict a dearth of BHs  with masses $\gtrsim 50\msun$, due to pulsational pair-instabilities affecting the progenitors of SBHs \citep[e.g.,][]{belc2012,fryer2012}. A natural way to produce MBHs in the (high-)mass gap would be though hierarchical mergers of smaller SBHs. If a MBH in the mass gap would then merge again with a SBH, the signal of this binary merger would appear again in the LIGO/Virgo GW frequency band \citep{abb16,abb17,kimb20}. This would require a dynamical environment with a high escape speed, where merger products could be retained. Globular clusters usually have lower escape speeds, thus most of the merger remnants would acquire a recoil kick large enough to overcome the cluster potential well \citep[e.g.][]{rodri2019}. As we showed, NSCs can have very high escape speeds and subsequent mergers can take place. Most of them can be detectable by  LIGO/Virgo, and for more massive BHs, by other instruments, such as Einstein Telescope (ET) or LISA.

The angular-averaged characteristic dimensionless strain amplitudes\footnote{That is, the GW strain in a logarithmic frequency bin or, equivalently, the Fourier transform of the time-domain dimensionless GW strain multiplied by the frequency. Angular averaging is over the binary orientation, sky position, and the detector’s antenna pattern.} of the GWs emitted by a source ($\mmbh$-$m_2$) in the inspiral phase at a luminosity distance $D$ is\footnote{Quantities $x_a$ are expressed with physical units $u$ as $x_{\mathrm{a},u} \equiv x_{\mathrm{a}}/u$, so that $x_{\mathrm{a},u}$ is dimensionless.} \citep[e.g.,][]{koc11}
\begin{align}\label{e:hc}
h_{\mathrm{c}}&\approx  
\begin{cases}
\frac{4}{5} \frac{\mathrm{G}^{5/3}}{\mathrm{c}^{4}}
\frac{M_{\mathrm{MBH}}^{5/3}m_{2} f_{\mathrm{GW}}^{7/6}}{D  }T_{\mathrm{obs}}^{1/2}
& {\rm if}~~f_{\mathrm{GW}}\leq f_{\mathrm{crit}}\,,\nonumber\\
\frac{1}{\sqrt{30}\pi^{2/3}}\frac{\mathrm{G}^{5/6}}{\mathrm{c}^{3/2}}\frac{M_{\mathrm{MBH}}^{1/3} m_{2}^{1/2} }{D f_{\mathrm{GW}}^{1/6}} & {\rm if}~~f_{\mathrm{GW}}\geq f_{\mathrm{crit}}\,,
\end{cases}
 \nonumber\\&=  
\begin{cases}
1.3\times10^{-23} M_{\mathrm{MBH,10^2M_{\odot}}}^{5/3} m_{\mathrm{2,10M_{\odot}}} 
f_{\mathrm{GW,10mHz}}^{\,5/6}T_{\mathrm{obs,yr}}^{1/2}
D_{\mathrm{Gpc}}^{-1}\\
\quad {\rm if}~~f_{\mathrm{GW}}\leq f_{\mathrm{crit}}\,,\\
6.5\times10^{-22} M_{\mathrm{MBH,10^2M_{\odot}}}^{1/3} m_{\mathrm{2,10M_{\odot}}}^{1/2} 
f_{\mathrm{GW,10Hz}}^{\,-1/6}
D_{\mathrm{Gpc}}^{-1}\\
\quad {\rm if}~~f_{\mathrm{GW}}\geq f_{\mathrm{crit}}\,, 
\end{cases}
\end{align}
where $f_{\rm crit}$ is given by the observation time $T_{\rm obs}$ as
\begin{equation}
f_{\rm crit} = 0.08\,M_{\mathrm{MBH,10^2M_{\odot}}}^{-1/4}
m_{\mathrm{2,10M_{\odot}}}^{-3/8}T_{\mathrm{obs,yr}}^{-3/8}\,\mathrm{Hz}\,.
\end{equation}
The typical frequency ($f_{\mathrm{GW}}$) for circular binaries is twice the orbital frequency, while there is a contribution from the eccentricity for eccentric inspirals \citep{wen2003}. If the observation time is much less than the inspiral time-scale and the source is circular, the source is approximately monochromatic with frequency $f_{\mathrm{GW}}\pm (1/2) T_{\mathrm{obs}}^{-1}$. For $f_{\mathrm{GW}}\lesssim f_{\rm crit}$, the GW frequency emitted by the binary is approximately constant during the observation time $T_{\mathrm{obs}}$, while, for $f_{\mathrm{GW}}\gtrsim f_{\rm crit}$, the binary inspirals during the observation and spans a frequency range up to the ISCO. Here the typical frequency becomes
\begin{equation}
f_{\mathrm{GW,ISCO}}=2f_{\mathrm{orb}}\approx 44 M_{\mathrm{MBH,10^2M_{\odot}}}^{-1}\mathrm{Hz}\,,
\end{equation}
for non-spinning IMBHs, while it is a factor $\sim 15$ higher for maximally spinning IMBHs. After the final inspiral phase, the merger and ring-down phases emit GWs at a higher characteristic ring-down frequency for zero spin
\begin{equation}
f_{\mathrm{RD}}\approx 120 M_{\mathrm{MBH,10^2M_{\odot}}}^{-1}\mathrm{Hz}.
\end{equation}
This is a factor $\sim 10$ higher for nearly maximal spins \citep{ber09}. 

\begin{figure*} 
\centering
\includegraphics[scale=0.55]{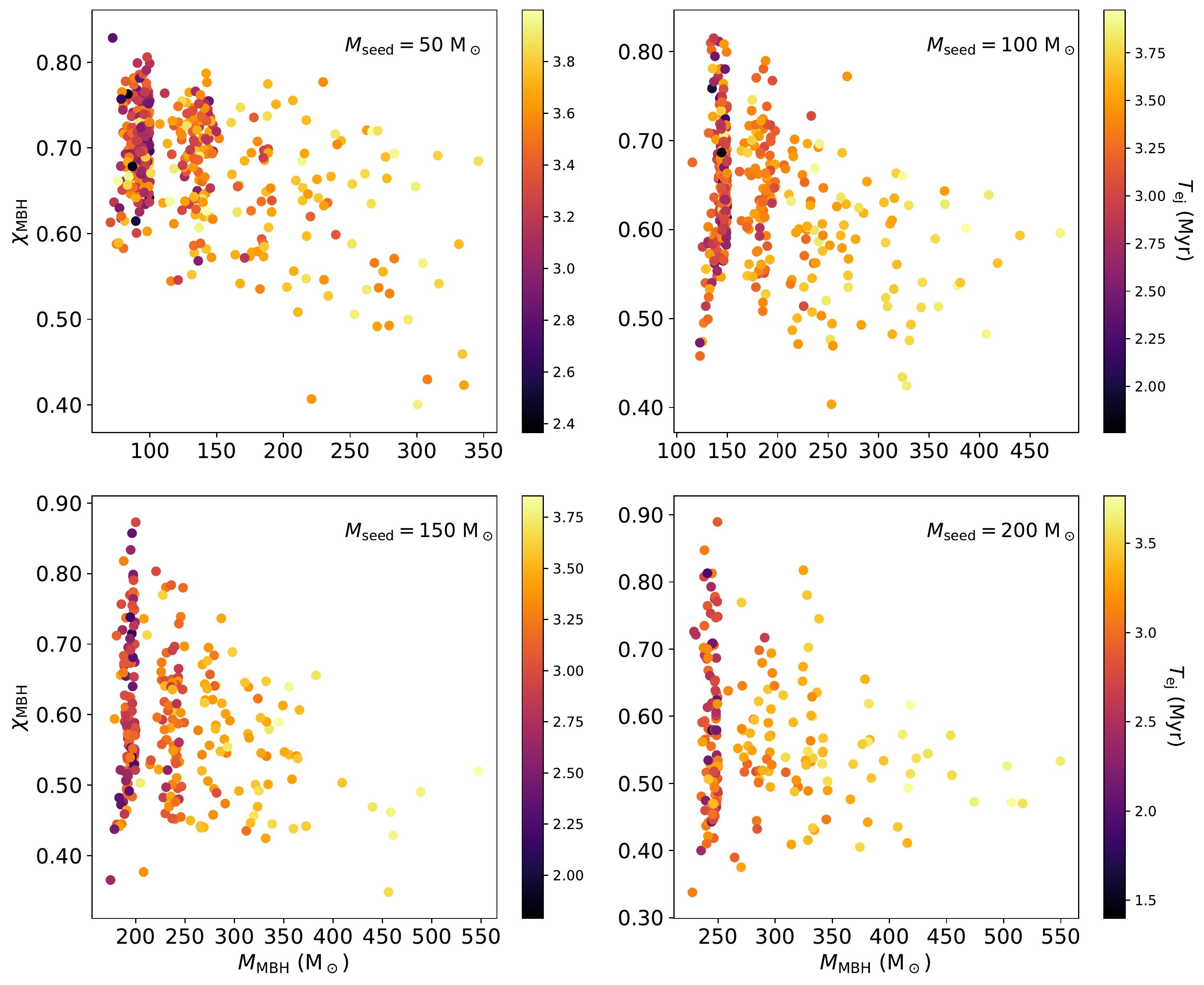}
\caption{Mass and spin of ejected MBH in a NSC of mass $M_{\rm NSC}=5\times 10^8\msun$ and density $\rho_{\rm NSC}=10^6\msun$ pc$^{-3}$. Seed mass: $50\msun$ (top-left panel); $100\msun$ (top-right panel); $150\msun$ (bottom-left panel); $200\msun$ (bottom-right panel). Color code: ejection time.}
\label{fig:eject1}
\end{figure*}

\begin{figure*} 
\centering
\includegraphics[scale=0.55]{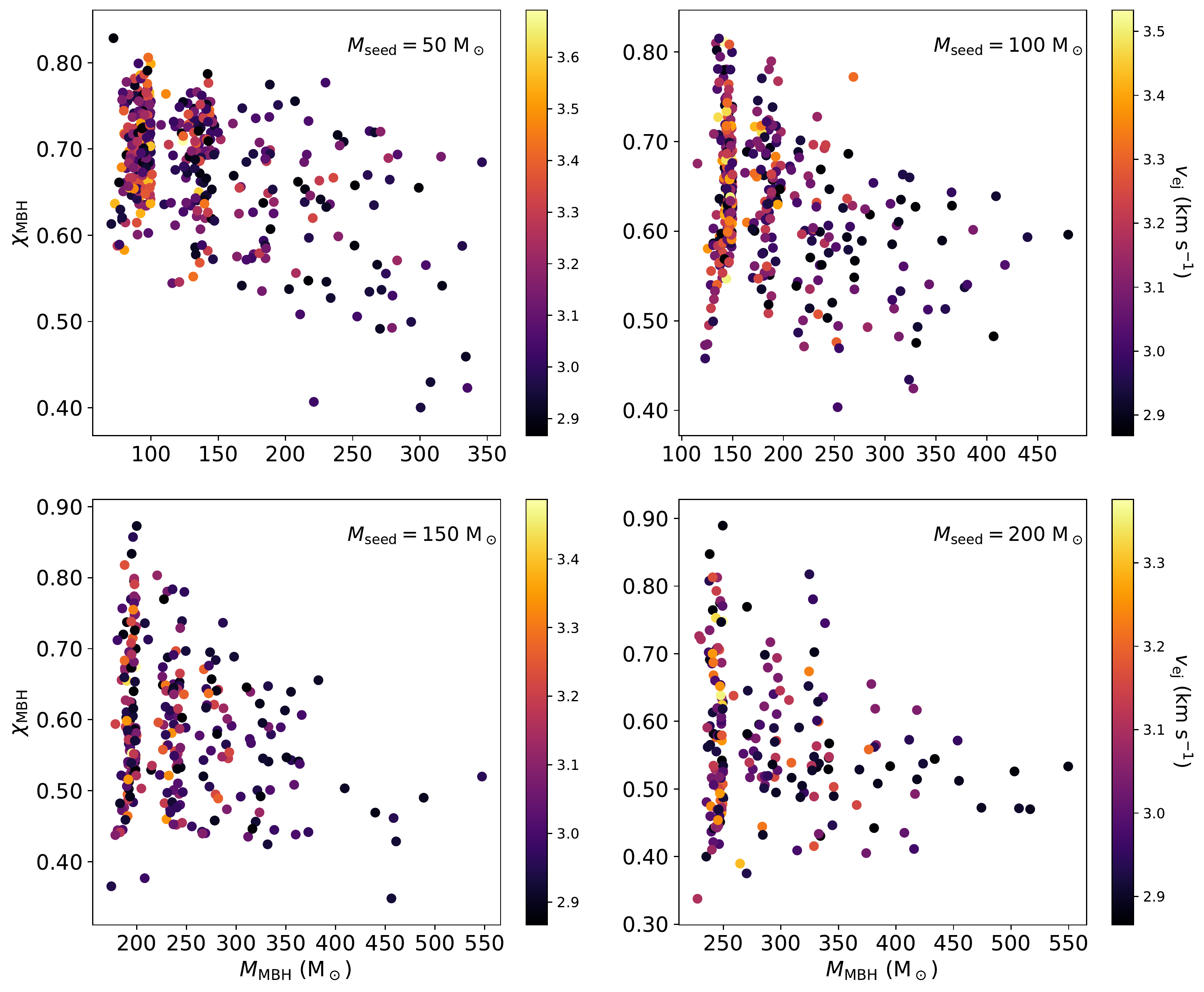}
\caption{Same as Figure~\ref{fig:eject1}. Color code: ejection velocity.}
\label{fig:eject2}
\end{figure*}

LIGO/Virgo is sensitive to GW frequencies in the range $\sim 10$--$10^3$ Hz. Mergers of MBHs in the pair instability mass gap and of IMBHs of a few hundred solar masses can therefore be observed by LIGO/Virgo. Some of them will first be observable by ET and LISA, and then by LIGO/Virgo. At the maximum luminosity distance attainable by  LIGO/Virgo, one  can observe an inspiral at a signal-to-noise ratio of $\sim 10$ is \footnote{Here masses are redshifted masses, which are the source-frame mass times $(1+z)$.}  \citep{fla98} 
\begin{equation}
 D_{\mathrm{LIGO,inspiral}}= 2.0\, 
 \left(\frac{M_{\mathrm{MBH}}}{100\mathrm{M}_{\odot}}\right)^{-1}
 \left(\frac{m_2}{10\mathrm{M}_{\odot}}\right)^{1/2}\,\mathrm{Gpc}\,.
\end{equation}

As an example, we report in Figure~\ref{fig:ligov} the distribution of mergers of MBHs for NSCs of mass $M_{\rm NSC}=5\times 10^6\msun$ (top-left), $M_{\rm NSC}=1\times 10^7\msun$ (top-right), $M_{\rm NSC}=1\times 10^8\msun$ (bottom-left), and $M_{\rm NSC}=5\times 10^8\msun$ (bottom-right). The NSC density is $10^6\msun$ pc$^{-3}$. The seed mass is $M_{\rm seed}=50\msun$, consistent with the lower boundary of the mass-gap \citep{belc2012,fryer2012}. Mergers in the mass gap can take place even in small NSCs, while the majority of them will occurs for masses $\gtrsim 1\times 10^7\msun$. For a detailed discussion on second-generation merges and MBHs in the mass gap see \citet{baibhav2020}.

ET and LISA are expected to be sensitive to GW frequencies between $\sim 0.1$--$10^4$ mHz, and $\sim 0.1$--$100$ mHz, respectively. As discussed above, some of the LIGO/Virgo mergers will first be observable by ET and LISA. On the other hand, mergers involving MBHs more massive than a few thousand solar masses will only merge in the LISA and/or ET band \citep[e.g.,][]{amaro2018}. The maximum luminosity distance at which these instruments can observe an inspiral (at a signal-to-noise ratio of $\sim 10$) is \citep{mil02a,gai11} 
\begin{align}
 D_{\mathrm{LISA,inspiral}}&= 1.9\, 
 \left(\frac{M_{\mathrm{MBH}}}{100\mathrm{M}_{\odot}}\right)^{1/2}
 \left(\frac{m_2}{10\mathrm{M}_{\odot}}\right)^{1/2}\,\mathrm{Gpc}\,,\\
  D_{\mathrm{ET,inspiral}}&= 10\, 
 \left(\frac{M_{\mathrm{MBH}}}{100\mathrm{M}_{\odot}}\right)^{1/2}
 \left(\frac{m_2}{10\mathrm{M}_{\odot}}\right)^{1/2}\,\mathrm{Gpc}\,.
\end{align}
Therefore, LISA and ET will offer a unique opportunity to observe and put constraints on the build-up of MBHs in NSCs.

\subsection{Ejected black holes}
\label{subsect:ejected}

As a consequence of the recoil kick, MBHs can be ejected even from massive and dense NSCs. The recoil kick velocity $\vkick$ is larger for higher spins of the merging BHs, thus the majority of the ejected MBHs are expected have rather high spins. At the same time, the recoil kick depends on the asymmetric mass ratio: the smaller the mass ratio, the smaller the kick (for non-zero spins). MBHs will not acquire a significant kick once they have grown to several hundreds solar masses and the mass ratio is $q\lesssim 0.1$. As a consequence, beyond an MBH mass threshold, merger remnants are preferentially retained within the NSC, rather than being ejected. Dynamical kicks are usually not high enough to eject MBHs from NSCs, unlike the case for globular clusters \citep{antoras2016}. 

We show in Figure~\ref{fig:eject1} the mass and spin of the ejected MBH in a NSC of mass $M_{\rm NSC}=5\times 10^8\msun$ and density $\rho_{\rm NSC}=10^6\msun$ pc$^{-3}$, and seed mass $50\msun$ (top-left panel), $100\msun$ (top-right panel), $150\msun$ (bottom-left panel), $200\msun$ (bottom-right panel). As expected, we find that the maximum mass of ejected MBH is $\sim 400$--$500\msun$, almost independently of the seed mass. More massive MBHs are retained, regardless of $M_{\rm seed}$. We find that MBHs are ejected at all times.

In Figure~\ref{fig:eject2}, we also illustrate the typical ejection speeds of the MBHs presented in Figure~\ref{fig:eject1}. We find that $\sim 86\%$, $\sim 76\%$, $\sim 55\%$, $\sim 40\%$ of the seeds are ejected for $M_{\rm seed}=50\msun$, $100\msun$, $150\msun$, $200\msun$. Moreover, we find that among the ejected MBHs $\sim 70\%$ and $\sim 19\%$ for $M_{\rm seed}=50\msun$, $\sim 64\%$ and $\sim 12\%$ for $M_{\rm seed}=100\msun$, $\sim 52\%$ and $\sim 5\%$ for $M_{\rm seed}=150\msun$, $\sim 46\%$ and $\sim 2\%$ for $M_{\rm seed}=200\msun$ have ejection velocities $>1000\kms$ and $>2000\kms$, respectively. For reference, the Milky Way's escape speed from the Galactic Center is $\sim 900$--$1000\kms$. Thus, a non-negligible fraction of ejected MBHs is ejected with kick velocities high enough to possibly escape the  galaxy. The fraction of these systems is smaller for higher seed masses and, obviously, for higher masses of the host galaxy, which in turn  correlates with the NSC mass \citep[e.g.,][]{georg2016}.

\subsection{Intermediate-mass black holes in dwarf galaxies}
\label{subsect:dwarfgal}

The end product of our merger scenario is the production of MBHs. Subsequent to their formation,  MBHs could  grow in NSCs by swallowing stars and by accreting gas to eventually become the  IMBHs that we observe today in the centers of dwarf galaxies \citep{korm2013,alex2017}. Such objects may eventually seed SMBHs in more massive galaxies by hierarchical merging, along with continued  infall of dwarf galaxies that contain  forming NSCs. This would  lead to merging of IMBH binaries as well as IMBH growth by tidal disruption and  Eddington-limited gas accretion. 

Although simplistic, our analysis has shown that in order to grow a seed  significantly, both the NSC mass and stellar density need to be sufficiently large. This comes from the fact that the growth is a compromise between the ability to retain the merger remnants and the rapidity in making binaries merge.

If this mechanism produces MBHs in galactic nuclei, dwarf galaxies could be naturally populated by IMBHs formed through repeated mergers. Such a mechanism overcomes the difficulty of accounting for  IMBH growth in dwarf galaxies  in the presence of SN feedback  \citep{2018MNRAS.478.5607T}.

\citet{silk2017} showed that current observations of active galactic nuclei in dwarfs could indeed be consistent with the  presence of IMBHs, provided that  the occupation fraction is  sufficiently high, in order   to  provide early feedback during the epoch of gas-rich galaxy formation. This could potentially yield a unifying explanation for many, if not all, of the dwarf galaxy \textquotedblleft anomalies\textquotedblright\ in a $\Lambda$CDM context. These  include the abundance, core-cusp, too-big-to-fail, ultra-faint, and baryon-fraction issues. 

One probe of this pathway to resolution of the many dwarf galaxy issues is that one would expect dwarf galaxies known to contain IMBHs to also display evidence of possibly relic nuclear star clusters. These are expected not just if the high central densities  density are  invoked to accelerate IMBH growth by MBH binary formation and merging,  but  are also required in order to accelerate the rate of tidal captures in the alternative  tidal disruption scenario for growing IMBHs \citep{2020arXiv200308133P}. 

There is indeed a correlation between the  presence of nuclear star clusters and  decreasing galaxy (and presumably IMBH) mass \citep{2020arXiv200103626N}. If dwarf galaxies were to generically  host IMBHs in their centers, there would be reduced   motivation for modifying the nature of cold dark matter in order to explain any of these dwarf galaxy issues.

\section{Discussion and conclusions}
\label{sect:conc}
The origin and the boundaries of the BH mass function are among the most puzzling questions in gravitational astrophysics. Even though LIGO/Virgo has detected several SBHs, the exact shape of the SBH mass spectrum remains a mystery since current stellar evolution models predict a dearth of BHs both with masses $\gtrsim 50\msun$ and $\lesssim 5\msun$ \citep{belc2017,fryer2012}. The dominant presence of IMBHs in dwarf galaxies  has not yet been demonstrated beyond any reasonable doubt, although there are several plausible examples  \citep{greene2019}.

A natural way to form both massive SBHs in the pair instability mass gap and  IMBHs is through repeated mergers of low-mass SBHs \citep{antonini2019}. To detect such objects through GW emission with current LIGO/Virgo or future LISA and ET, the merger remnant has to acquire a new companion with which to merge. Moreover, the host environment has to be massive and dense enough to have a large escape speed in order to retain the merger remnant, which acquires a recoil kick as a result of anisotropic GW emission.

In this paper, we have explored the possibility that MBHs may be born as a result of repeated mergers of a seed of mass $M_{\rm seed}$ with low-mass SBHs in NSCs. We have shown  how the typical distribution of MBH masses depends on the NSC mass and density. Remarkably, we have found that the MBH growth is a delicate balance between the ability of a NSC to retain the merger remnants when they acquire a recoil kick and the rapidity at which the retained MBHs can form and merge in new binaries. Massive NSCs can more easily retain  MBHs, but the formation of binaries that merge takes place on longer time-scales. On the other hand, denser NSCs both retain  MBHs more easily and more efficiently form binaries that merge through GW emission. We have also explored the role of the initial spin, and found that low initial spins lead to the production of a larger population of MBHs. We have also discussed how mergers of SBHs in the mass gap as well as  IMBH can be observed by LIGO/Virgo, ET, and LISA. Finally, we have shown that the mass of the typical ejected MBH is $400$--$500\msun$. These objects are ejected with velocities of upto a few thousands $\kms$, enough in some cases to escape their host galaxy.

We caution that, even though our current model gives the general pathways for the formation of MBHs in NSCs, it still lacks  many details that are relevant to place NSC into a broader context. First of all, we have neglected a detailed evolution of the NSC based on a star-to-star basis. Unfortunately, current codes can only handle up to $\sim 10^7$ particles \citep[e.g.,][]{gier2006MNRAS.371..484G,Pattabiraman2013}. Secondly, NSCs are not isolated systems, but they have a continuous supply of stars and compact objects from the rest of the galaxy \citep{alex2017}. Thirdly, whenever a massive object forms, it  occupies the innermost central region of a NSC and creates a cusp of stars and compact objects \citep[e.g.,][]{bahc1976,hopale2006,frasar2018}. Fourth, the growing seed can also accreate gas, rendering our current estimates a lower limit. This depends on many factors and, ultimately, on the details of the complex history of the host NSC \citep[e.g.][]{roupas2019,kroupa2020}. Finally, we note that episodic star formation and accretion of star clusters can lead to morphological and structural transformation of the nuclei which is difficult to address with our simplified models \citep{ant2014}. We leave the detailed exploration of each of these limitations to future work.\\
\ \\
\textit{Data Availability}\\
\ \\
The data underlying this article will be shared on reasonable request to the corresponding author.\\

\section*{Acknowledgements}

We thank Fabio Antonini for insightful comments. GF acknowledges support from a CIERA postdoctoral fellowship at Northwestern University. GF acknowledges hospitality from the Johns Hopkins University, where this work was initiated.

\bibliographystyle{mn2e}
\bibliography{refs}

\end{document}